# Quantum Critical Behaviour in the Superfluid Density of Strongly Underdoped Ultrathin Cuprate Films

Iulian Hetel, Thomas R. Lemberger[*], and Mohit Randeria

Department of Physics, The Ohio State University, Columbus, Ohio 43210, USA

[*]e-mail: TRL@mps.ohio-state.edu

**A central issue in the physics of high temperature superconductors is to understand superconductivity within a single copper-oxide layer or bilayer, the fundamental structural unit in the cuprates, and how it is lost with underdoping. As mobile holes are removed from the $CuO_2$ planes, the transition temperature $T_C$ and superfluid density $n_S$ decrease in a surprisingly correlated fashion in crystals and thick films[1-4]. We seek to elucidate the intrinsic physics of bilayers in the strongly underdoped regime, near the critical doping level where superconductivity disappears. We report measurements of $n_S(T)$ in films of $Y_{1-x}Ca_xBa_2Cu_3O_{7-\delta}$ as thin as two copper-oxide bilayers with $T_C$'s as low as 3 K. In addition to seeing the two-dimensional (2D) Kosterlitz-Thouless-Berezinski transition[5,6] at $T_C$, we observe a remarkable scaling of $T_C$ with $n_S(0)$ that demonstrates that the disappearance of superconductivity with underdoping is due to *quantum* fluctuations near a T = 0 2D quantum critical point.**

Early measurements[1] of the suppression of superfluid density $n_S$ and $T_C$ for moderately underdoped cuprates prompted the suggestion that *thermal* fluctuations[7] of the phase of the superconducting order parameter were the primary cause. This interpretation neatly accounted for observations of approximately linear scaling[1]: $T_C \propto n_S(0)$, and a wide critical region[8] near $T_C$. It was widely accepted that these behaviours would persist all the way to the disappearance of superconductivity with underdoping. Thus, it was a surprise when recent measurements on strongly underdoped $YBa_2Cu_3O_{7-\delta}$ crystals[2,3] and films[4] showed that the scaling of $T_C$ with $n_S(0)$ is actually sublinear: $T_C \propto [n_S(0)]^\alpha$ with $\alpha \approx$ ½. Moreover, the critical region was much smaller than in moderately underdoped samples[9]. These observations motivated the new hypothesis[10,11] that underdoping leads to the disappearance of superconductivity at a 3D quantum critical point (QCP), as opposed to a first-order quantum phase transition. In this paper we show that the behaviour of ultrathin films is consistent with a 2D quantum critical point. It is difficult to see how any theory other than quantum criticality could account for the observed scaling, its dependence on dimensionality, and insensitivity to disorder.

It has taken a long time to produce persuasive studies of the superfluid density of strongly underdoped cuprate superconductors because it is difficult to produce sufficiently homogenous samples. We were able to do so in $YBa_2Cu_3O_{7-\delta}$ (YBCO) films by reducing the oxygen concentration in the CuO chain layers of this compound nearly to zero, i.e., overall oxygen stoichiometry, 7-$\delta \approx$ 6, thereby removing holes but also reducing inhomogeneity arising from the CuO chains. We compensated by doping holes into the $CuO_2$ planes with the replacement of 20 to 30% of $Y^{3+}$ with $Ca^{2+}$. We grew $Y_{0.8}Ca_{0.2}Ba_2Cu_3O_{7-\delta}$ and $Y_{0.7}Ca_{0.3}Ba_2Cu_3O_{7-\delta}$ (Ca-YBCO) films as thin as 2 unit cells (1 uc = 1.17 nm) by pulsed laser deposition onto atomically flat $SrTiO_3$ substrates, with a calibrated growth rate of 17 pulses per unit cell thickness. X-ray measurements show that our films are epitaxial, with the highly conducting $CuO_2$ layers parallel to the substrate.

We probe the superfluid of $CuO_2$ layers by using a two-coil mutual-inductance method[12,13] at a frequency $\omega/2\pi$ = 50 kHz. This method provides the sheet conductivity, $Y \equiv (\sigma_1 + i\sigma_2)d$, of our superconducting films, where $d$ is the film thickness, and $\sigma_1 + i\sigma_2$ is the usual complex conductivity. Precautions are taken to ensure that data are taken in the linear-response regime, *i.e.*, Y is independent of the size of the 50 kHz magnetic field produced by the drive coil. The superfluid density $n_S$ is proportional to the nondissipative part, $Y_2 = \sigma_2 d$, and is therefore closely related to the magnetic penetration depth $\lambda$ *via* the relation, $n_S(T) \propto \mu_0 \omega Y_2/d \equiv 1/\lambda^2(T)$. In the following, we refer to $1/\lambda^2$ as the superfluid density.

The highest $T_C$ that we achieved in our 2 uc Ca-YBCO films was about 52 K, comparable to the maximum $T_C$ that has been observed[14-16] in 2 uc YBCO films without Ca. For these highly-doped 2 uc films we observed $1/\lambda^2(0) \approx 14$ $\mu m^{-2}$, which is comparable to values obtained in thick YBCO films and 40 uc thick Ca-YBCO films with comparable $T_C$'s, (see below). On this basis, we assert that thin and thick films have similar structural and stoichiometrical quality. Comparison of ultrathin films, thick films, and ultraclean crystals, presented below, further support this assertion. $1/\lambda^2$ and the real conductivity, $\sigma_1$, for the most underdoped 2-uc thick films are plotted vs. T in Fig. 1a. The films show a single, reasonably narrow, peak in $\sigma_1$. In our experience, the fact that $\sigma_1$ returns to zero below the transition is a reliable indicator of good homogeneity. All samples reported here have this feature.

We now discuss the nature of the finite temperature phase transition, and then turn to the scaling of $T_C$ with superfluid density. The key qualitative feature in $1/\lambda^2$ is its abrupt downturn as T increases. Figure 1b shows $1/\lambda^2(T)$ for representative underdoped 2-uc Ca-YBCO films over a wide range of doping. The top 3 curves (brown) represent measurements on the same film, right after its growth (highest $T_C$), and after it lost oxygen while sitting at room temperature. Similarly, the next 8 curves (blue) represent the same film at different oxygen concentrations. All other curves represent different as-grown samples that had been sealed with an amorphous cap layer to eliminate oxygen loss. As seen in Fig. 1, while there are some sample-to-sample variations in the details, all samples show the same basic features, namely, $1/\lambda^2$ is flat, (approximately quadratic), at low T and has an abrupt downturn as T increases.

The Kosterlitz-Thouless-Berezinski (KTB) theory of thermally-excited vortex-antivortex (V-aV) pairs in 2D superconducting films predicts a super-to-normal phase transition marked by a discontinuous drop in superfluid density[6]. The transition temperature $T_C$ is the temperature where $1/\lambda^2(T) = 8\pi\mu_0 k_B T/d\Phi_0^2$. ($d$ = film thickness and $\Phi_0 = 2\pi\hbar/2e$ is the flux quantum.) This relationship strictly applies only if $1/\lambda^2$ is measured at zero frequency. The right-hand-side of this relationship is plotted as a dashed line in Figs.1a and 1b. Its intersection with $1/\lambda^2$ measured at 50 kHz, not *dc*, approximates the predicted $T_C$. In the simplest scenario, we would expect the intersection to occur at the onset of the downturn in $1/\lambda^2$, as it does in 2D films of superfluid helium-4[17,18] measured at 5 kHz. Instead, it consistently occurs closer to the middle of the drop in $1/\lambda^2$. Given the complexities of cuprate films, e.g., grain boundaries, vortex pinning, residual inhomogeneity, and the likelihood of new physics (e.g., ref. 19), none of which is included in the KTB theory, we do not expect the KTB theory to fit our data quantitatively. The point that we wish to draw from Fig. 1 is that, regardless of details, 2 uc thick films with a wide range of doping levels are consistent among themselves, and their behaviours point to a 2D super-to-normal transition mediated by unbinding of V-aV pairs.

In Fig. 2 we show our results for $T_C$ vs. $1/\lambda^2(0)$ for thick and thin films. We first note that data on ultrathin films (red circles) at moderate underdoping ($T_C$'s from 20 to 50 K) overlap data on thick films, as noted above. Thick (20 – 40 uc) YBCO films[4] (black circles) and our thick (40 uc) Ca-YBCO films (green circles) agree quantitatively with each other at all dopings, and both show the scaling: $T_C \propto [1/\lambda^2(0)]^\alpha$, where $\alpha \approx 0.5$, (c.f. dashed line in Fig. 2). We emphasize that this scaling is insensitive to disorder in our films, because high-purity YBCO single crystals (orange and blue squares) exhibit the same scaling[2,3], despite the fact that their superfluid densities at T = 0 are several times larger than those of films. The most important part of Fig. 2 is at strong underdoping, where a striking difference between the 2D and 3D samples emerges. For ultrathin films, $T_C$ drops more rapidly with underdoping, and the relationship between $T_C$ and $1/\lambda^2(0)$ is close to linear: $T_C \propto [1/\lambda^2(0)]^\alpha$, where $\alpha \approx 1$ (c.f., solid lines in Fig. 2). As a consequence, their superfluid densities exceed not only the values measured in thick films, but also those of clean YBCO crystals with similar $T_C$'s.

To understand the difference in scaling for 2D and 3D samples, let us look at predictions of theory assuming that underdoping destroys superconductivity at a quantum critical point (QCP).[20] Quite generally, an energy scale like $T_C$ vanishes as one approaches the QCP like $T_C \propto \delta^{z\nu}$, where $\delta$ measures the deviation in doping from the QCP, and z and ν are the dynamical and the correlation length exponents, respectively.

The precise values of these universal critical exponents will turn out to be unimportant for our purposes. Josephson scaling[21] near a QCP implies that the T=0 superfluid density vanishes as $n_S(0) \propto \delta^{(z+D-2)\nu}$ where D is the spatial dimensionality. We may eliminate $\delta$ between these two relations to obtain the scaling relationship[10,11] $T_C \propto n_S(0)^{z/(z+D-2)}$ between the two quantities measured in our experiment.

In D = 3 dimensions, theory finds $T_C \propto n_S(0)^{z/(z+1)}$. Because the exponent z is expected to be between 1 and 2 on general grounds, we expect $T_C \propto n_S(0)^\alpha$, where $\alpha$ is between 1/2 and 2/3. This prediction is consistent with all of the data on 3D samples. Turning now to the ultrathin films, in D = 2 dimensions, theory predicts linearity, $T_C \propto n_S(0)$, independent of the value of z. On this basis, we conclude that quantum phase fluctuations near a 2D QCP are responsible for the linear scaling that we observe.

To put our results into context, we note that there are several competing ordered states in the severely underdoped region of the cuprate phase diagram, including d-wave superconductivity, antiferromagnetism of the undoped Mott insulator and, possibly, charge ordering. Under these conditions, one might generically expect to observe a first order phase transition from superconductivity to some other ordered state, rather than a quantum critical point, at the doping where superconductivity disappears. It is thus quite remarkable that our 2D and 3D data, taken together, strongly support the presence of a T=0 quantum critical point at the superconductor-to-nonsuperconductor transition with underdoping.

**Methods.**

**Sample preparation.**

Ultrathin $Y_{1-x}Ca_xBa_2Cu_3O_{7-\delta}$ films with their copper-oxide planes parallel to the substrate are grown by pulsed laser deposition (140 mtorr $O_2$, heater temperature 760 C, energy density 2 J/cm$^2$, growth rate 0.70 Å/pulse) on $SrTiO_3$ substrates. Thin insulating layers of $PrBa_2Cu_3O_{7-\delta}$ protect the film above and below. By replacing up to 30% of $Y^{3+}$ with $Ca^{2+}$, we need only a small concentration of oxygen to obtain superconductivity, especially in the strongly underdoped region, and we achieve a wide range of hole doping with greatly reduced inhomogeneity in the CuO chains. The oxygen concentration in the films was controlled by the oxygen pressure during growth and cool-down. To minimize oxygen loss at room temperature, on some samples an amorphous $PrBa_2Cu_3O_{7-\delta}$ layer was deposited at a temperature, 280-300C, too low for the perovskite structure to form. It is possible that some Ca diffuses into adjacent $PrBa_2Cu_3O_{7-\delta}$ layers during the growing process, and that the actual Ca concentration is lower than its nominal value. In order to obtain reproducible growth and continuous ultrathin superconducting layers, atomically flat substrate surfaces were prepared by controlled buffered-HF etching [22,23], and checked by atomic force microscopy.

**Acknowledgements**. We are grateful to Zlatko Tesanovic who greatly encouraged this investigation and provided many useful comments. We thank David Stroud, C. Jayaprakash and Y. Zuev for many useful discussions. This work was supported in part by NSF-DMR grant 0203739. IH is grateful to The Ohio State University for an OSU Presidential Fellowship.

Figure 1 **Superfluid density vs. T for 2 unit cell thick $Y_{1-x}Ca_xBa_2Cu_3O_{7-\delta}$ films**. Superfluid density $n_S$ is proportional to the inverse square magnetic penetration depth, $n_S \propto 1/\lambda^2$. Intersection of dashed lines with $1/\lambda^2(T)$ is approximately where a 2D transition is predicted[6]: $1/\lambda^2(T) = 8\pi\mu_0 kT/d\Phi_0^2$, where $d$ = film thickness. **a**. $1/\lambda^2(T)$ (left axis) and real conductivity $\sigma_1(T)$ (right axis) for the two most strongly underdoped ultrathin films. The T dependence of $\sigma_1$ is an indicator of good film homogeneity. **b**. $1/\lambda^2(T)$ for the full range of doping that we studied. Top three (brown) curves are the same film in three conditions: as-grown (highest $T_C$), and with two lower dopings, after some oxygen diffused out of the film at room temperature. Similarly, the next 8 (blue) curves represent the same film at different doping levels. Additional curves each represent different as-grown films.

Figure 2 **Scaling of $T_C$ with superfluid density at T = 0.** $T_C$ vs. $1/\lambda^2(0)$ on a log-log scale for our Ca-YBCO films 2 unit cells thick (red dots) and 40 uc thick (green dots). $T_C$ is defined from the midpoint of the drop in $1/\lambda^2$, and error bars extend from the top of the drop to where $1/\lambda^2$ is 5% of its value at T = 0. For reference we include Uemura's µSR results on YBCO powders[1] (open black circles), $H_{c1}$ measurements[2] on clean YBCO crystals (open orange squares), microwave measurements[3] on clean YBCO crystals (open blue squares), and data on 20-40 uc YBCO films[4] (black dots). Solid lines illustrate the linear relationship, $T_C \propto 1/\lambda^2(0)$, that describes our strongly underdoped ultrathin films and is expected near a 2D quantum critical point. The dashed line illustrates a square root relationship, $T_C \propto \sqrt{1/\lambda^2(0)}$, that describes strongly underdoped 3D samples (crystals and thick films) and is consistent with 3D quantum criticality.

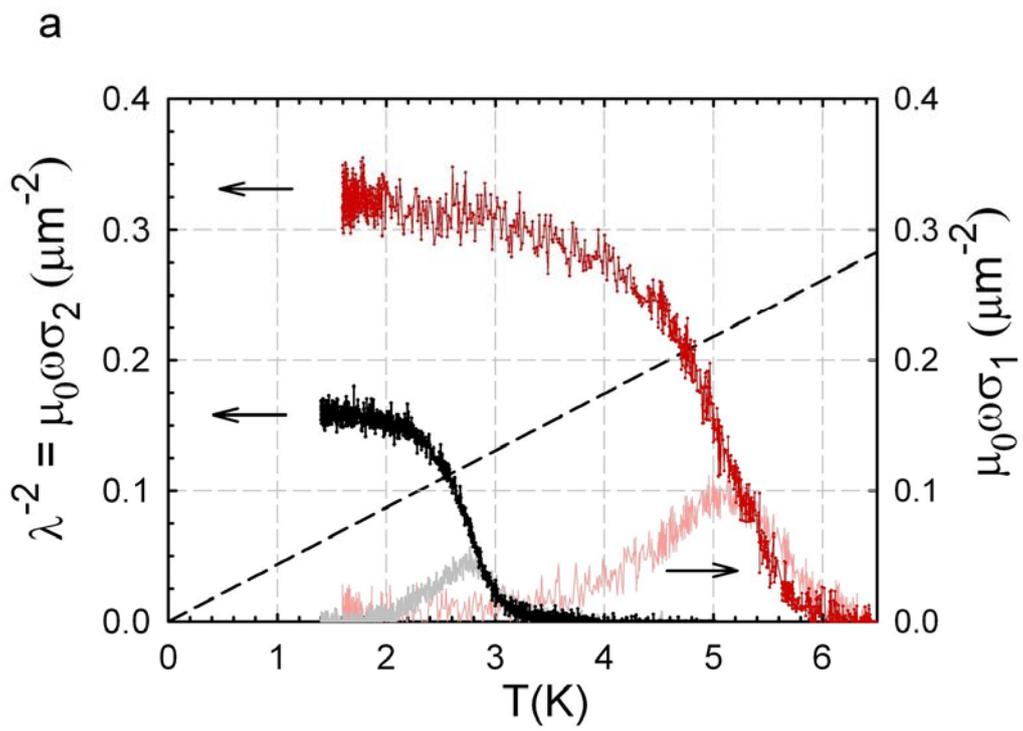

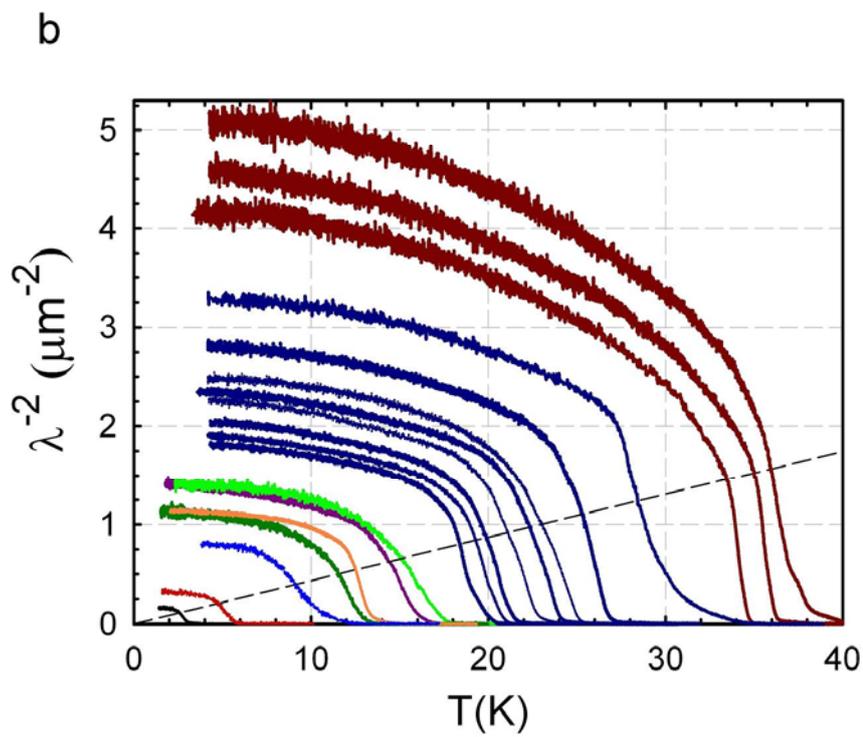

**Figure 1**

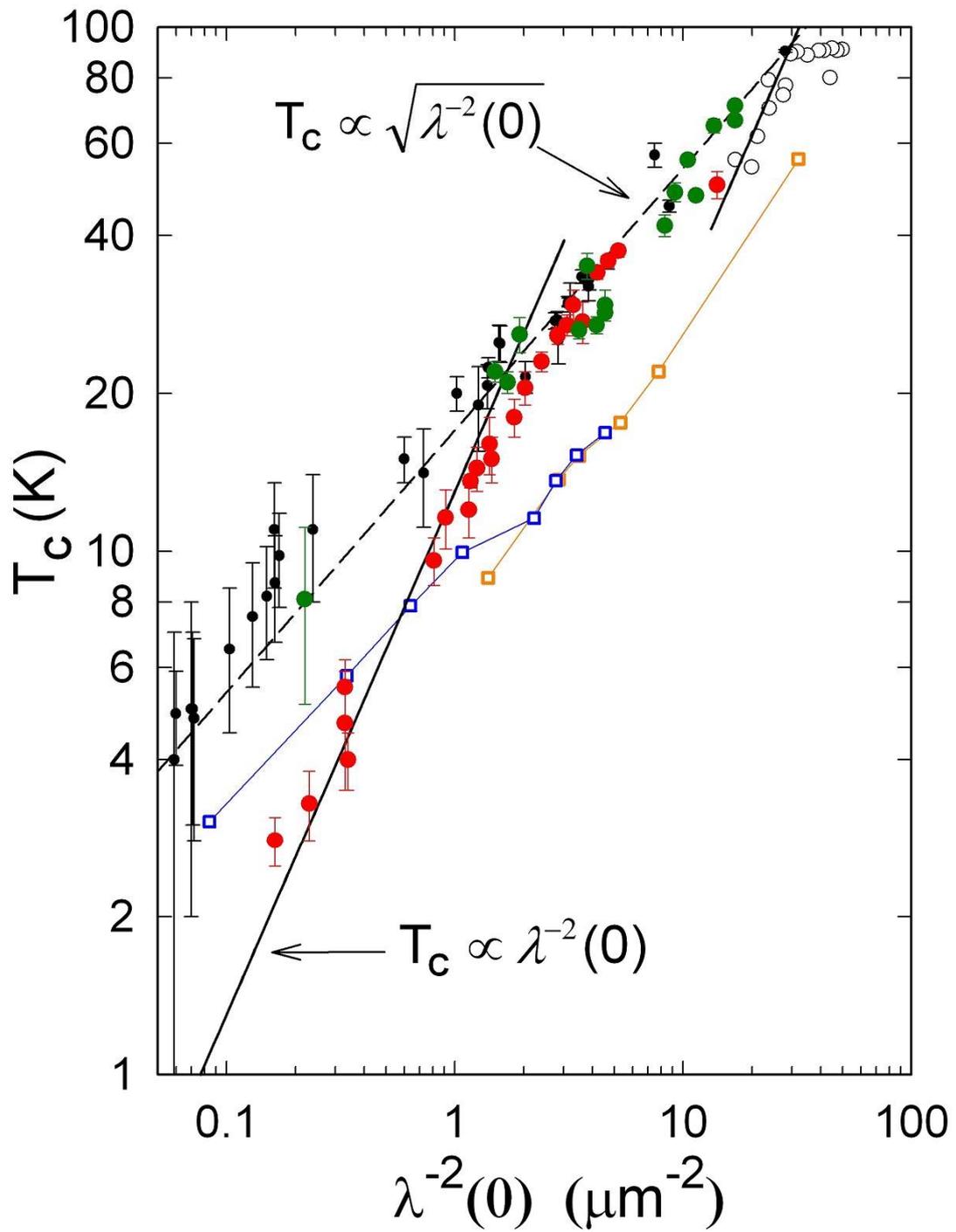

**Figure 2**